\documentclass[pdflatex,sn-mathphys]{sn-jnl}

\jyear{2022}%

\usepackage{lineno}

\begin{document}

\title[]{Onset of Bloch oscillations in the almost-strong-field regime}

\author[1]{\fnm{Jan} \sur{Reislöhner}}

\author[1]{\fnm{Doyeong} \sur{Kim}}

\author[2,3,4]{\fnm{Ihar} \sur{Babushkin}}

\author[1]{\fnm{Adrian N.} \sur{Pfeiffer}}

\affil[1]{\orgdiv{Institute of Optics and Quantum Electronics, Abbe Center of Photonics}, \orgname{Friedrich Schiller University}, \orgaddress{\street{Max-Wien-Platz 1}, \postcode{07743} \city{Jena}, \country{Germany}}}

\affil[2]{\orgdiv{Institute for Quantum Optics}, \orgname{Leibniz Universität Hannover}, \orgaddress{\street{Welfengarten 1}, \postcode{30167} \city{Hannover}, \country{Germany}}}

\affil[3]{\orgdiv{Cluster of Excellence PhoenixD (Photonics, Optics, and Engineering – Innovation Across Disciplines)}, \orgaddress{\street{Welfengarten 1}, \postcode{30167} \city{Hannover}, \country{Germany}}}

\affil[4]{\orgdiv{Max Born Institute}, \orgaddress{\street{Max Born Str. 2a}, \postcode{12489} \city{Berlin}, \country{Germany}}}





\maketitle

\textbf{
In the field of high-order harmonic generation from solids, the electron motion typically exceeds the edge of the first Brillouin zone. In conventional nonlinear optics, on the other hand, the excursion of band electrons is negligible. 
Here, the transition from conventional nonlinear optics to the regime where the crystal electrons begin to explore the first Brillouin zone is investigated. It is found that the nonlinear optical response changes abruptly already before intraband currents due to ionization become dominant. This is observed by an interference structure in the third-order harmonic generation of few-cycle pulses in a non-collinear geometry. Although approaching Keldysh parameter $\gamma = 1$, this is not a strong-field effect in the original sense, because the iterative series still converges and reproduces the interference structure. The change of the nonlinear interband response is attributed to Bloch motion of the reversible (or transient or virtual) population, similar to the Bloch motion of the irreversible (or real) population which affects the intraband currents that have been observed in high-order harmonic generation.  
}

A crystal electron accelerated by an electric field in an electronic band (a Bloch electron) shows a motion pattern that is distinctively different from a free electron. 
The motion of a Bloch electron is described in $k$-space by the acceleration theorem $\partial_t k = -E$ \cite{RN218} (atomic units are used). 
Already Felix Bloch recognized that the electron motion under the influence of a constant electric field would be oscillatory instead of unidirectional, because the group velocity $v_{n}^k = \partial_k \omega_{n}^k $ of an electron wave packet in band $n$ with energy $\omega_{n}$ flips the sign after crossing the Brillouin zone edges. 
However, these Bloch oscillations, one of the most intriguing and counter-intuitive corollaries of electronic bands, are difficult to observe because electron scattering prevents extended motion in $k$-space for static fields below the breakdown threshold.

One option to realize Bloch oscillations is to increase the lattice constant $a$ so that the zone edge at $k=\pi/a$ is closer. This was achieved by semiconductor superlatices and allowed the first observations of Bloch oscillations \cite{RN250,RN251,RN252}. Another option is to increase the electric field sufficiently so that the zone edge is reached ultrafast before scattering destroys the wave packet. This condition can be met for intense laser pulses when the electric field is so strong that electrons cross the zone edge within one optical cycle. This was considered from the beginning as a possible mechanism of high-order harmonic generation (HHG) from transparent crystals \cite{RN255}. For most conditions, the contribution of Bloch oscillations to HHG was reported to be weaker than other mechanisms \cite{RN158}, such as recollision \cite{RN241}, multiband coupling \cite{RN261}, and motion in bands with higher spatial frequencies \cite{RN262}. For terahertz fields it was found using a numerical switch-off analysis that Bloch oscillations contribute substantially to HHG \cite{RN293}. Very recently, it has been reported that HHG produced by two-color fields exhibit phase variations that can be associated with reaching the zone edges \cite{RN294}. On the other hand, the decoherence dynamics of strong-field processes in solids are disputed, with many recent calculations assuming ultrafast loss of interband coherences with few-femtosecond dephasing times \cite{RN241, RN224, RN242, RN221, RN239}. If the underlying reason of this ultrafast coherence loss is rooted in scattering, it is questionable how laser-driven Bloch oscillations can arise.

\begin{figure}[h]%
\centering
\includegraphics[width=1\textwidth]{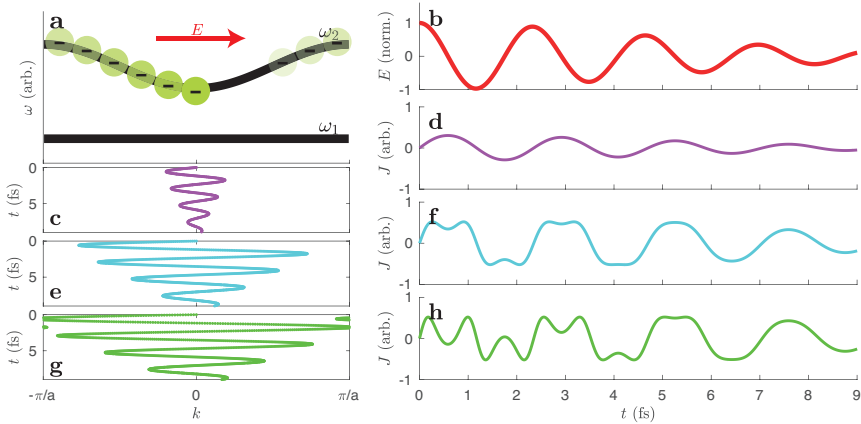}
\caption{\textbf{Current generated by Bloch electrons.} \textbf{a}, An electron that is born in the conduction band at $k=0$  will subsequently undergo motion according to the acceleration theorem. With the electric field displayed in \textbf{b}, the electron trajectories for peak intensity $I = $ 1\,TW/cm$^2$, 15\,TW/cm$^2$ and 30\,TW/cm$^2$ are depicted in \textbf{c, e, g}. The corresponding currents are shown in \textbf{d, f, h}. Parameters: lattice constant $a = 0.5$\,nm; peak electric field $\max (E) = \sqrt{\frac{8 \times 10^{-7} \pi c}{n_R}} I $ ; refractive index $n_R = 1.54$.}
\label{figSing}
\end{figure}

For low-order harmonics, it is commonly accepted that ionization gains importance as the damage threshold is approached \cite{RN243}. Recently, the influence of the step-wise ionization on intraband currents has been discussed \cite{RN135, RN265, RN295}. When the electrons begin to cover a significant range of the first Brillouin zone, the contribution of Bloch electrons to low-order harmonics is expected to change, as visualized in Fig.\,\ref{figSing}. An electron promoted to the conduction band ($n=2$) generates the current $J(t) = -v_{2}^{k(t)}$. With a rather short-range trajectory at low intensities, the current contains mainly fundamental frequencies. With increasing intensity, the electron motion becomes anharmonic and contains increasingly more third harmonic frequencies. 

Here, this simple picture is extended to the motion of interband coherences, which give rise to the interband polarization. It is found that the nonlinear optical response from interband coherences (which are responsible for the reversible or transient or virtual population) changes abruptly as the crystal electrons explore the first Brillouin zone. This is observed by an interference structure in the third-order harmonic generation (THG). For short laser pulses, this happens already at intensities where the contribution of intraband currents due to ionization is not yet dominant. The mechanism is different from the influence of Bloch electron motion on intraband currents (which are seeded by the irreversible or real population), which has been observed before in HHG \cite{RN293, RN294}. The observed effect is in the realm of perturbative optics, because the iterative series \cite{RN249} converges and reproduces the interference structure. The regime of intensities might be called almost-strong-field regime, because the intensities are smaller than in strong-field laser physics, yet the nonlinear response differs substantially from conventional nonlinear optics. Conventional nonlinear optics is understood here to indicate that the excursion of band electrons is negligible. Strong-field laser physics, on the other hand, is understood here to indicate that the response cannot be treated perturbatively, which implies a Keldysh parameter $\gamma < 1$. The almost-strong-field regime, where the electron trajectories cover a significant range of the first Brillouin zone but the contribution of intraband currents is still negligible, is commonly reached in lenses and windows of high-power optical instruments, in contrast to the strong-field regime, where optical elements deteriorate quickly. 

\begin{figure}[h]%
\centering
\includegraphics[width=1\textwidth]{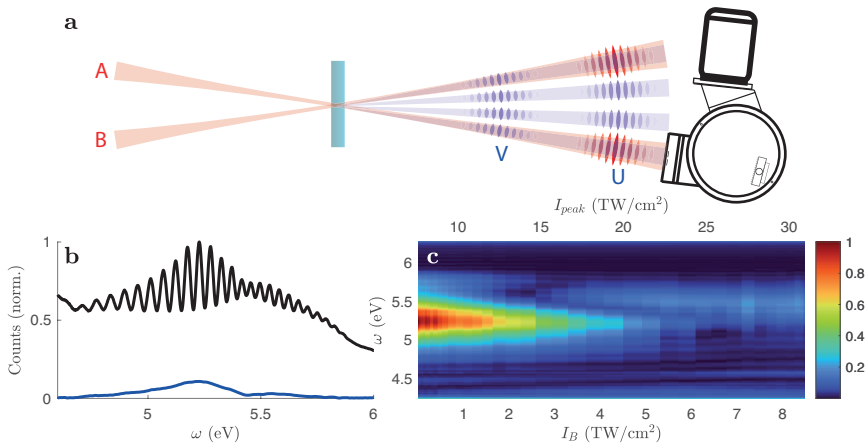}
\caption{\textbf{Experiment.} \textbf{a}, Two fundamental beams (red) are focused into a crystal (crossing angle $\alpha = 1.1^{\circ}$, beam waist 120\,$\mu$m). Beams in the DUV (blue) are emitted from the laser-induced grating both collinearly to the fundamental beams and and in the interstitial space. The wavefronts of the fundamental pulses A and B, which originate by beam splitting of one laser pulse, are highlighted in red. Cross-phase modulation scans (XPMS) are used for experimental pulse retrieval \cite{RN192} and indicate a center wavelength of 700\,nm and a pulse duration of 8\,fs. The DUV beams contain two pulses (wavefronts highlighted in blue), of which the front pulses travel at the pace of the fundamental pulses, whereas the rear pulses propagate at a speed corresponding to the DUV dispersion and are thus delayed due to the refractive index mismatch at Vis-IR and DUV wavelengths. The spectrometer records the DUV spectrum emitted collinearly to A. The experiment is carried out in vacuum, inhibiting nonlinear interactions with air. 
\textbf{b}, The raw spectrum $I(\omega)$ (black) and the Fourier filtered spectrum $I_r(\omega)$ (blue) at $I_{B}$ = 3.5\,TW/cm$^2$ from SiO$_2 (001)$. 
\textbf{c}, The Fourier filtered spectrum $I_r(\omega)$ as a function of intensity.}
\label{figExp}
\end{figure}

To uncover the expected interference, an intensity scan is required that extends from the regime of conventional nonlinear optics into the regime of Bloch electron motion. This introduces significant complications for the experiment, as the yield of harmonic light would cover a very wide range. To overcome this problem, non-collinear spectroscopy is used here, sketched in Fig.\,\ref{figExp}. Two visible-infrared (Vis-IR) pulses A and B are focused into 100-$\mu$m-thick crystals with polarization perpendicular to the plane of incidence. A and B are overlapped spatially and temporally with a precision of $\pm 10\mu $\,m and $\pm 1$\,fs, forming a laser induced grating \cite{RN18}. Deep ultraviolet (DUV) light is produced by THG in the crystals. The DUV light that is emitted collinearly to A, which is kept at a constant intensity $I_{A}$ = 7\,TW/cm$^2$, is detected with a spectrometer. The intensity of B is varied in the range $I_{B}$ = [0, 8]\,TW/cm$^2$, which varies the peak intensity $I_{peak} = \left( \sqrt{I_{A}} + \sqrt{I_{B}} \right)^2 $ in the grating. This facilitates intensity scans over a high dynamic range with little variation in spectrometer count rates.

The fundamental Vis-IR pulses are strongly modified by nonlinear pulse propagation, which obscures the observation of THG mechanisms. Fortunately, THG from the beginning of the crystals strongly contributes to the fringed DUV spectrum (Fig.\,\ref{figExp}\,\textbf{b}). As previous studies \cite{RN192, RN194} revealed, the spectral fringes are produced by two DUV pulses that are well separated in time after the crystal. The leading pulse (labeled U) is in time with the Vis-IR pulse, whereas the trailing pulse (labeled V) propagates at the DUV group velocity \cite{RN192, RN194, RN263}. V is beneficial for the interpretation of the data, because it is generated within the first few micrometers and maintains its spectrum in the subsequent linear propagation. The contribution of U complicates the interpretation of the data, because it is generated after nonlinear pulse propagation modified the fundamental pulses. Due to scattering of the Vis-IR light in the DUV spectrometer, the spectra contain also a significant background in addition to the fringed spectra $I \left( \omega \right) = \left\vert U\left( \omega \right) + V\left( \omega \right) \right\vert^2$. To remove the background and to enhance the sensitivity to V, the raw spectra are inverse Fourier transformed, the side peak (alternating component) is cut-out and shifted to zero, and thereafter Fourier transformed. This yields $I_r \left( \omega \right) = \left\vert U^*(\omega)V(\omega) \mathrm{e}^{-i \omega t_e} \right\vert$, where $t_e$ represents the shift to zero which corresponds to the delay between U and V after the medium. The parameters used are 87\,fs for SiO$_2$, 111\,fs for Al$_2$O$_3$, 190\,fs for MgO. 

\begin{figure}[h]%
\centering
\includegraphics[width=1\textwidth]{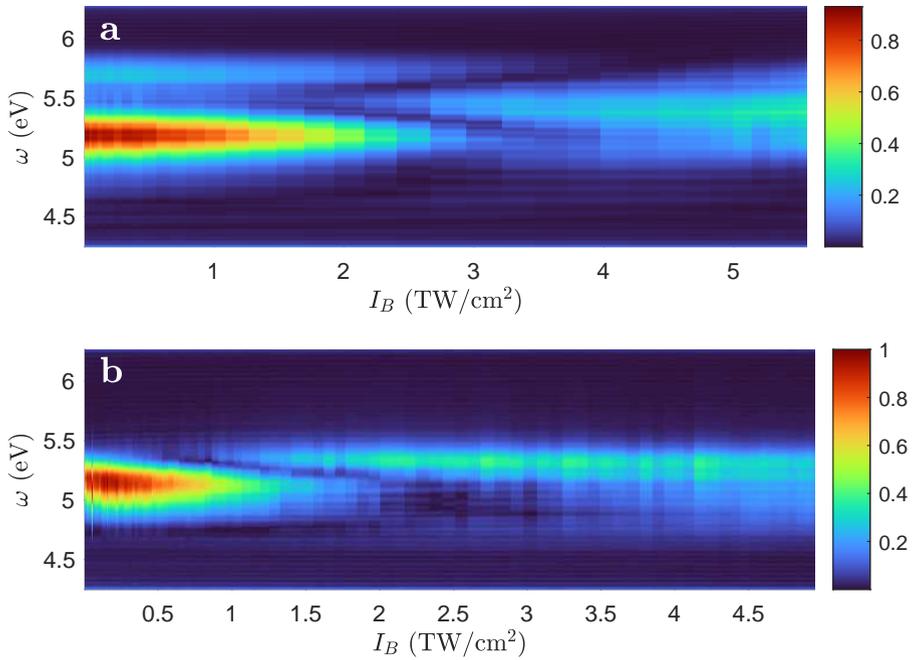}
\caption{\textbf{(Extended Data) Comparison with other crystals}. The Fourier filtered spectrum $I_r(\omega)$ as a function of intensity as in Fig.\,\ref{figExp}, but for Al$_2$O$_3 (110)$ (\textbf{a}) and MgO $(100)$ (\textbf{b}).}\label{figExpComp}
\end{figure}

The intensity scan reveals an interference structure in the region $I_{peak}$ = [10, 20]\,TW/cm$^2$ for SiO$_2$, see Fig.\,\ref{figExp}\,\textbf{c}. As this is the regime where the crystal electrons explore the first Brillouin zone (Fig.\,\ref{figSing}), this is a first indication for THG from Bloch electrons that competes with conventional THG. Corresponding experiments in Al$_2$O$_3$ and MgO yield similar interferences (Extended Data Fig.\,\ref{figExpComp}), but fine details indicate that the band structure has an influence. 

To get further insights, numerical calculations are performed using semiconductor Bloch equations (SBEs) \cite{RN218}. With restriction to the spatial dimension of the electric field vector and omitting the Coulomb interaction, the SBEs read \cite{RN249, RN222}
\begin{align}
i \frac{d}{dt} \rho_{nm}^{k+A} = -\omega_{nm}^{k+A} \rho_{nm}^{k+A}
+ E \cdot \sum_{l} \left( d_{lm}^{k+A} \rho_{nl}^{k+A} - d_{nl}^{k+A} \rho_{lm}^{k+A} \right) \nonumber\\
+ i \left( \partial_t  \rho_{nm}^{k+A} \right)_{relax}
. \label{VelGauge} 
\end{align}

The diagonal elements $\rho_{nn}^k$ of the density matrix are the populations of the electronic bands, the off-diagonal elements $\rho_{nm}^k (n \neq m)$ are the coherences between the states with transition energies $\omega_{nm}^k = \omega_{m}^k - \omega_{n}^k$. The electric field $E$ induces dynamics by coupling the electronic bands via the dipole matrix elements $d_{nm}^k$ and by moving the electrons and holes within the bands, which is realized by the coordinate transform $k \to k + A$, where $A$ is the vector potential defined by $E = - \partial_t A$. The relaxation terms $\left( \partial_t \rho_{nm}^k  \right)_{relax}$ are implemented as phenomenological damping terms (see Methods). The polarization $P$ and the current $J$ are calculated by 
\begin{align}
P &= \sum_{n \ne m} \sum_k d_{nm}  \rho_{nm}^k \delta k \label{equ:P} 
\\
J &= - \sum_{n} \sum_k \rho_{nn}^k  v_{n}^k \delta k
, \label{equ:J} 
\end{align}
where $\delta k$ is the spacing in the $k$-grid. 

In most recent studies, the band structures are calculated by \textit{ab initio} methods like density functional theory (DFT). Here, a different approach is taken. For a quantitative comparison of optical fields originating from macroscopic pulse propagation, it is essential that both the linear response (including the group velocity dispersion) and the nonlinear response (including the optical Kerr effect (OKE)) match the experiment. The linear response of DFT is known to deviate because of missing background contributions \cite{RN257}; if the OKE is correctly reproduced is typically not tested. Here, numerical refractive index data is used to incorporate the linear polarization in the pulse propagation. The nonlinear polarization $P^{\mathrm{(NL)}}$ is used as a source term in the pulse propagation, which is calculated from equ.\,\ref{equ:P} in Fourier space by
\begin{equation}
\mathcal{F} \left\{P^{\mathrm{(NL)}}  \right\}
= \mathcal{F} \left\{ P \right\} - \chi^{(1)} \mathcal{F} \left\{ E \right\} ,
\end{equation}
where $\chi^{(1)}$ is the linear susceptibility of equ.\,\ref{VelGauge} (see Methods). The dipole matrix elements of three-bands are then adjusted to match experimental data of the OKE. This supports pulse propagation using the unidirectional pulse propagation equation (UPPE) at feasible computation times, yet captures the interband and intraband dynamics consistently \cite{RN249}. 

\begin{figure}[h]%
\centering
\includegraphics[width=1\textwidth]{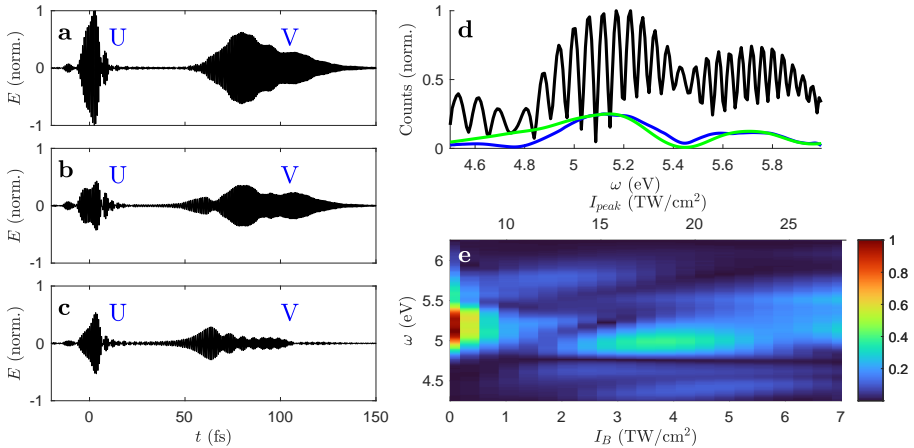}
\caption{\textbf{Macroscopic calculations using experimentally determined pulse shapes in SiO$_2$.} The electric field of the DUV pulses U and V emitted collinearly to A for $I_{peak} = 7$\,TW/cm$^2$ (\textbf{a}), 14\,TW/cm$^2$ (\textbf{b}) and 23\,TW/cm$^2$ (\textbf{c}). The electric field was Fourier filtered to include only frequencies in the interval [4.2, 6.2]\,eV to suppress the much stronger fundamental field at Vis-IR wavelengths.
\textbf{d}, The raw spectrum $I(\omega)$ (black) and the Fourier filtered spectrum $I_r(\omega)$ (blue) at $I_{peak}$ = 14\,TW/cm$^2$. The spectrum $\mathcal{F} \left\{P^{\mathrm{(NL)}} \right\}$ calculated with the initial fundamental pulses at the beginning of the crystal is shown in green for comparison. 
\textbf{e}, The Fourier filtered spectrum $I_r(\omega)$ as a function of intensity.} 
\label{figProp}
\end{figure}

The macroscopic UPPE calculations (see Methods), using the experimentally determined pulse shapes, confirm the assumption of two separate DUV pulses, see Fig.\,\ref{figProp}. The calculated spectra are processed like the experimental spectra. The Fourier-filtered spectrum $I_r(\omega)$, although somewhat masked by propagation effects, still resembles the THG at the beginning of the crystal (Fig.\,\ref{figProp}(\textbf{d})). The reason is that V originates within the first few micrometers in the crystal and maintains its spectrum in the subsequent linear propagation. The intensity scan exhibits an interference structure that is similar to the experimental data. In order to investigate the influence of the band shape, calculations are performed where the band shape contains higher frequencies (Extended Data Fig.\,\ref{figBand3}). As expected, the band shape influences the interference structure, which might be exploited to extract information about band shapes from the data. However, the calculations limited to three bands are unlikely to reproduce the interference structure in fine detail.

The interference structure also address a debated inconsistency in the field of HHG from solids. Numerical calculations by several groups agree that noisy spectra of HHG are predicted, in contrast to the clean harmonics measured experimentally. The most prominent solution of this discrepancy is to assume ultrafast coherence loss realized by dephasing times below $10$\,fs, which helps the calculations produce clean harmonics \cite{RN241, RN224, RN242, RN221, RN239}. The interference structure vanishes for such short dephasing times (Extended Data Fig.\,\ref{figTdeph}), which does not support the assumption of dephasing times below 10\,fs. 

\begin{figure}[h]%
\centering
\includegraphics[width=0.7\textwidth]{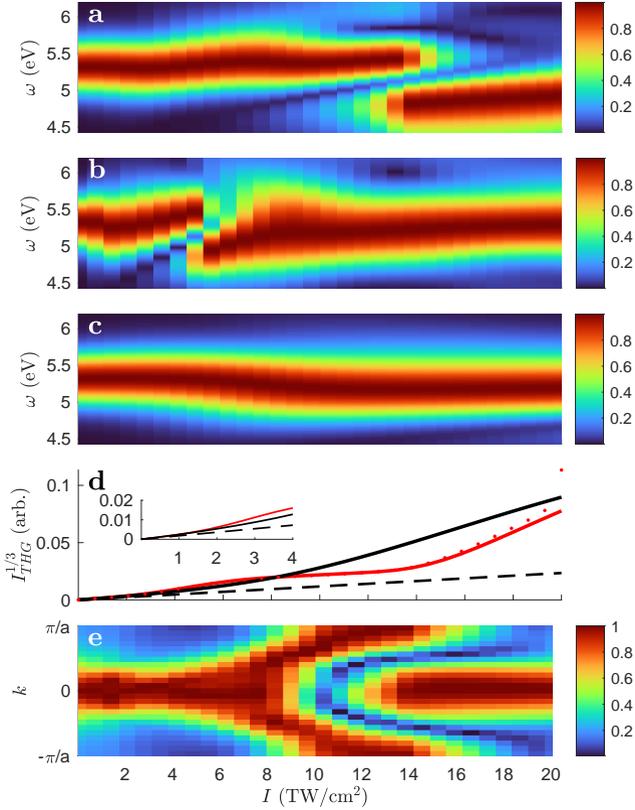}
\caption{\textbf{Optically thin calculations using 8-fs Gaussian pulses.} The spectra of $P^{\mathrm{(NL)}}$ (\textbf{a}) and $J$ (\textbf{b}) are calculated using the full SBEs. For comparison, the spectrum of $P^{\mathrm{(NL)}}$ with suppression of the electron motion is shown in (\textbf{c}). All pseudo color plots are normalized at each intensity to increase the visibility. The third root of the sum of the spectra of $P^{\mathrm{(NL)}}$ before normalization (corresponding to the third root of the total THG intensity) are depicted in (\textbf{d}) for the SBE calculation with (red solid) and without (black solid) Bloch electron motion. The red dotted line shows the iterative calculation with 100 iterations that diverges for $I> 18$\,TW/cm$^2$. For comparison, the black dashed line shows the instantaneous response $P^{\mathrm{(NL)}} = \chi^{(3)} E^3$, which is a straight line in this diagram. The $k$-resolved contribution $P^{\mathrm{(NL)}}(k)$ of the full-SBE calculation summed over frequencies [4.5, 5]\,eV is displayed in \textbf{e}.
}
\label{figPandJ}
\end{figure}

\begin{figure}[h]%
\centering
\includegraphics[width=1\textwidth]{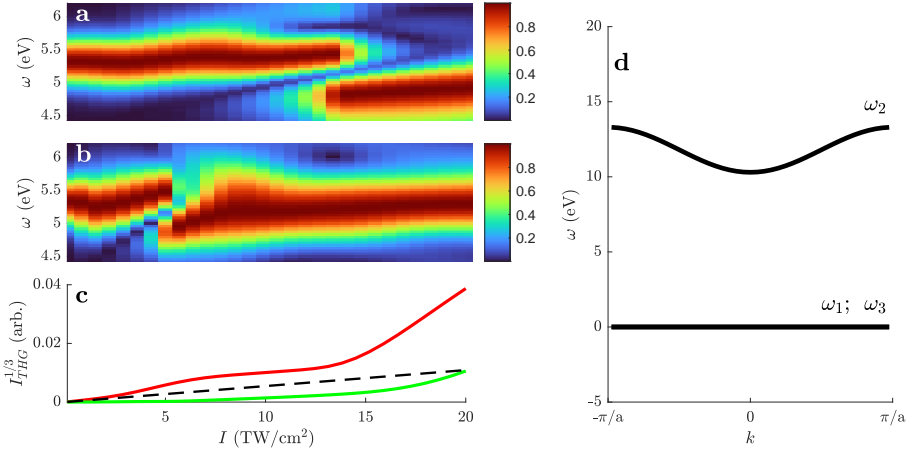}
\caption{\textbf{(Extended Data) The influence of numerical approximations.} Time-domain integrations of (\ref{VelGauge}) using 8-fs Gaussian pulses in a three band model are performed. Unlike in the main text (Fig.\,\ref{figPandJ}), the same approximations as for the iterative series ($\rho_{11}^k -\rho_{22}^k = 1$; $\rho_{11}^k -\rho_{33}^k = 0$; $\rho_{33}^k -\rho_{22}^k = 1$) are used. The spectra of $P^{\mathrm{(NL)}}$ and $J$ are shown in \textbf{a} and \textbf{b}. The spectra have been normalized at each intensity to increase the visibility. The third root of the sum of $P^{\mathrm{(NL)}}$ (red solid) and $J / 3 \omega_{12}$ (green solid) before normalization are depicted in (\textbf{c}), corresponding to the contribution of the nonlinear polarization and the current to the total THG intensity. For comparison, the black dashed line shows the instantaneous response $P^{\mathrm{(NL)}} = \chi^{(3)} E^3$, which is a straight line in this diagram. The spectra and curves are almost identical to those in Fig.\,\ref{figPandJ}, justifying the approximations of the iterative series. The band structure used is shown in \textbf{d}. }
\label{figSBE_o}
\end{figure}

\begin{figure}[h]%
\centering
\includegraphics[width=1\textwidth]{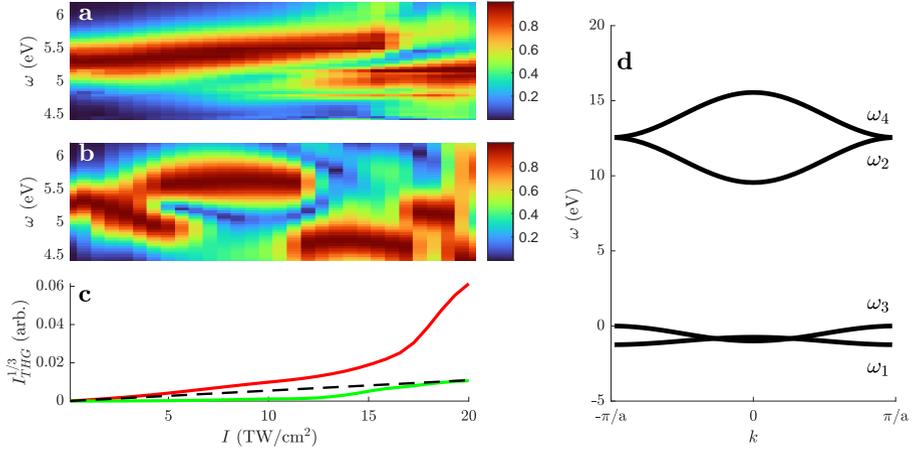}
\caption{\textbf{(Extended Data) Comparison to a more realistic band structure.} The same as (Extended Data) Fig.\,\ref{figSBE_o} but using a more realistic band structure. Four bands are used as shown in \textbf{d}. The band energies are $\omega_n^k = \omega_n + \frac{1}{2} b_n (1 - \mathrm{cos}(ka))$ with $\omega_n = -1$\,eV; 9.3\,eV; -1.25\,eV; 15.3\,eV and $b_n = -0.5$\,eV; 3\,eV; 1\,eV; -3\,eV. The valence to conduction band transitions $d_{12}^{k} = d_{21}^{k}$ are implemented as in the main text (\ref{DipoleElement}). The valence band transitions are implemented with $d_{13}^k = d_{31}^k = 12$. The conduction band transitions are implemented with $d_{24}^k = d_{42}^k = 12$. All other dipole matrix elements are set to zero. This band structure reproduces the same OKE as the band structure used in (Extended Data) Fig.\,\ref{figSBE_o} and is very similar to the structure used in Ref.\,\cite{RN163}.
}
\label{figSBE_Schiff}
\end{figure}

\begin{figure}[h]%
\centering
\includegraphics[width=1\textwidth]{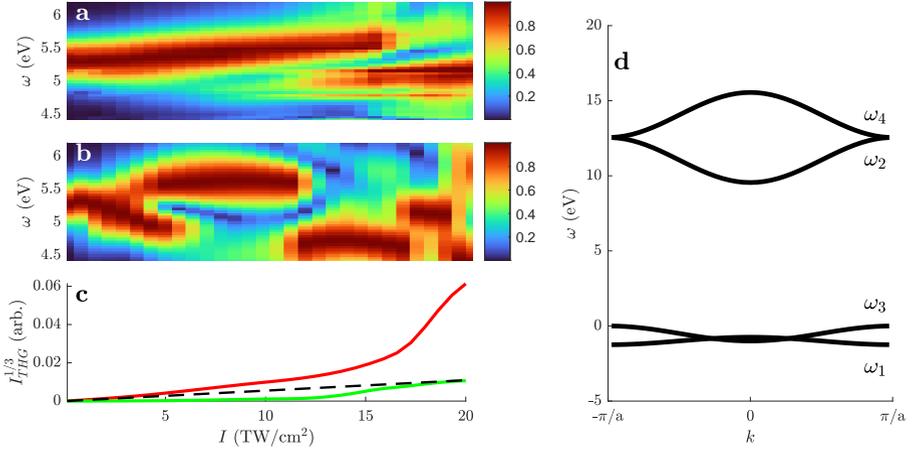}
\caption{\textbf{(Extended Data) Comparison to an asymmetric band structure.} The same as (Extended Data) Fig.\,\ref{figSBE_Schiff} but using an asymmetric band structure. The conduction band energy is $\omega_2^k = \omega_2 + \frac{1}{2} b (1 - \mathrm{cos}(ka))) - c \mathrm{sin}(ka)$ with $\omega_2 = 9.3$\,eV, $b = 3$\,eV and $c = 1$\,eV.
}
\label{figSBE_assym}
\end{figure}

To clarify whether the interference structure is caused by a change of the mechanism of THG rather than propagation effects, the nonlinear response generated by 8-fs Gaussian pulses is investigated, see Fig.\,\ref{figPandJ}. $P^{\mathrm{(NL)}}$ exhibits an interference structure in the range [8, 20]\,TW/cm$^2$, similar to the UPPE calculations and the experiment. The shape is influenced by the band shape (see Extended Data Figs.\,\ref{figSBE_o}, \ref{figSBE_Schiff}, \ref{figSBE_assym}), but the general appearance is universal. Also $J$ shows an interference structure, but at lower intensities [2, 8]\,TW/cm$^2$. This is an indication that the interference structure observed experimentally is not due to $J$. This is affirmed by running the UPPE calculation with $J=0$, which yields an indistinguishable result from the complete calculation displayed in Fig.\,\ref{figProp}\,\textbf{c}. This seems to contradict recent wave-mixing experiments at similar intensities \cite{RN135, RN265}, but a crucial point may be that the SBE calculations reproduce the reversible population of the conduction band \cite{RN134, RN182, RN249} rather than assuming a step-wise ionization.

The Keldysh parameter $\gamma = \frac{\omega_0 \sqrt{\omega_{12}}}{ \left\vert E \right\vert }$, where $\omega_0$ is the optical frequency, is commonly used to distinguish multiphoton ($\gamma > 1$) and strong-field ($\gamma<1$) interactions \cite{RN226}. While the former can be treated by the power series expansion of perturbative nonlinear optics, this series diverges for the latter.
The transition region has attracted much attention for gases, where it is sometimes referred to as the regime of nonadiabatic tunneling \cite{RN254}, but has not yet received much attention for solids. The intensity range of the interference structure ($ \gamma = 1.5$ at 8\,TW/cm$^2$ and $\gamma = 1$ at 18\,TW/cm$^2$) is below the regime of strong-field laser physics in the original sense. Strong-field laser physics in the original sense is understood here to mean that the power series expansion does not converge. To test the convergence, the SBEs are solved iteratively (see Methods and \cite{RN249}). It has been shown before that the convergence criterion of the iteration is fulfilled for $\gamma > 1$ \cite{RN249}. The pseudo color spectra produced with 100 iterations are distinguishable from those of the time-domain integration displayed in see Fig.\,\ref{figPandJ}\,\textbf{a}. Only the line plot of Fig.\,\ref{figPandJ}\,\textbf{d} reveals deviations starting at 18\,TW/cm$^2$ where $\gamma = 1$. 

To finally reveal the mechanism that causes the modification of the non-linear response in the almost-strong-field regime, a simplified calculation is performed neglecting the motion of the Bloch electrons. This is achieved by omitting the coordinate transform $k \to k + A$ in (\ref{VelGauge}). With the Bloch electron motion turned off, the interference structure in $P^{\mathrm{(NL)}}$ disappears. For lower intensities, displayed in the inset of Fig.\,\ref{figPandJ}\,\textbf{d}, the calculations with and without Bloch electron motion agree perfectly. Thus, Bloch electron motion can be neglected at moderate intensities. Moreover, the instantaneous response $P^{\mathrm{(NL)}} = \chi^{(3)} E^3$, which is a common simplification for the OKE in transparent solids, is a very good approximation for these intensities. In the almost-strong-field regime, the THG intensity deviates from the instantaneous response model for both calculations with and without Bloch electron motion but only the full calculation generates the interference structures. At 5\,TW/cm$^2$, where $J$ exhibits interference, the electrons transverse up to 45\% of the first Brillouin zone. However, this is not observed in the experiment because the influence of $J$ is still negligible at these low intensities. At 14\,TW/cm$^2$, where $P^{\mathrm{(NL)}}$ exhibits interference, the electrons transverse up to 75\% of the first Brillouin zone. The origin of the nonlinear polarization in $k$-space is traced by omitting the $k$-summation in (\ref{equ:P}) resulting $P^{\mathrm{(NL)}}(k)$. In the almost-strong-field regime, the origin is shifted through the entire Brillouin zone (Fig.\,\ref{figPandJ}\,\textbf{d}). This underpins the interpretation of Bloch motion that affects the interband polarization. At low intensities, only the local band curvature is decisive which is highest at $k=0$. When the electrons start to explore the Brillouin zone, the band curvature throughout the trajectory must be considered both for the interband and for the intraband contribution of the nonlinear response. 

\begin{figure}[h]%
\centering
\includegraphics[width=1\textwidth]{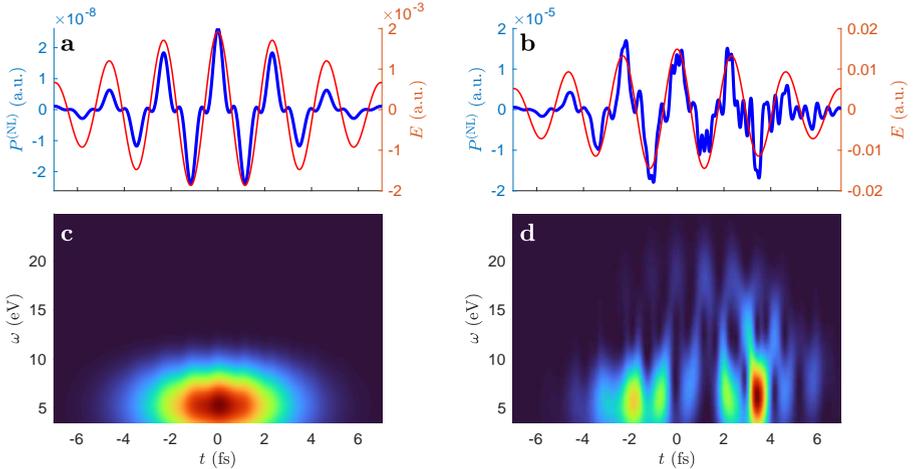}
\caption{\textbf{(Extended Data) The nonlinear polarization response in time-domain.} $P^{\mathrm{(NL)}}$ is calculated using the SBEs with Bloch electron motion at 0.2\,TW/cm$^2$ (\textbf{a} and \textbf{c}) and at 4\,TW/cm$^2$ (\textbf{b} and \textbf{d}). The blue lines in  \textbf{a} and \textbf{b} show $P^{\mathrm{(NL)}}$, the generating field is displayed by the red curve. The spectrograms (\textbf{c} and \textbf{d}) show the smoothed pseudo Wigner-Ville distributions of $P^{\mathrm{(NL)}}$. Before calculating the spectrograms, all frequency components of $P^{\mathrm{(NL)}}$ with $\omega < 3.5$\,eV were removed by Fourier filtering.
}
\label{figSpecP}
\end{figure}

\begin{figure}[h]%
\centering
\includegraphics[width=1\textwidth]{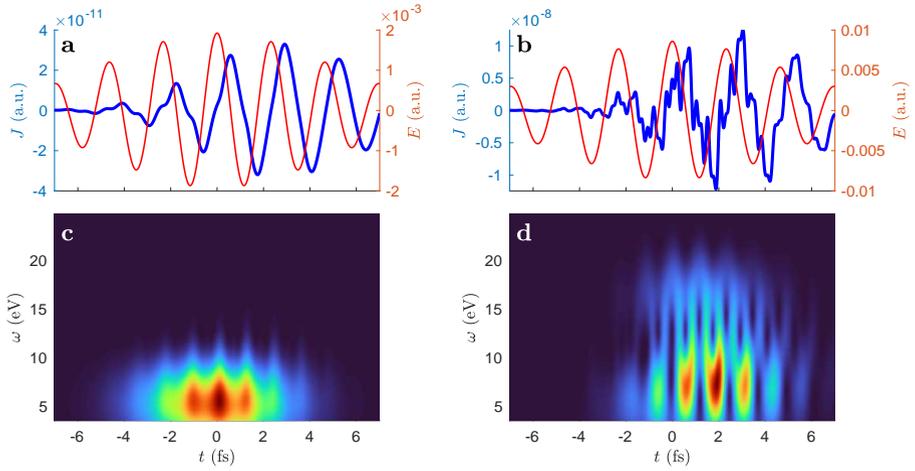}
\caption{\textbf{(Extended Data) The current in time-domain}. The same as Fig.\,\ref{figSpecP}, but $J$ instead of $P^{\mathrm{(NL)}}$ is displayed.}
\label{figSpecJ}
\end{figure}

\begin{figure}[h]%
\centering
\includegraphics[width=1\textwidth]{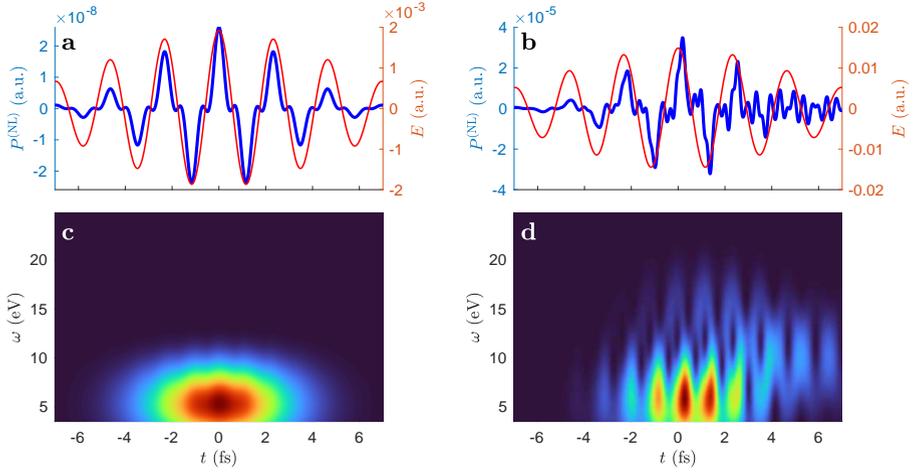}
\caption{\textbf{(Extended Data) The nonlinear polarization response without Bloch electron motion}. The same as Fig.\,\ref{figSpecP}, but using the simplified calculation that neglects the motion of the Bloch electrons by omitting the coordinate transform $k \to k + A$ in (\ref{VelGauge}). The peaks of $P^{\mathrm{(NL)}}$ in \textbf{b} are not dented as compared to Fig.\,\ref{figSpecP}}
\label{figSpecPnm}
\end{figure}

The spectrograms (Fig.\,\ref{figSpecP} and Extended Data Fig.\,\ref{figSpecJ}) show that while THG is temporally delocalized at low intensity, THG and also higher frequency generation are localized within the optical cycle at higher intensities. Some features are reminiscent of the three-step model for HHG \cite{RN258}. In particular, there are branches with positive chirp, as typically associated with short electron trajectories, followed by branches with negative chirp, as typically associated with long trajectories. However, the three-step model would predict only photons with energies of band transitions \cite{RN241, RN261}, which are limited to [10.3, 13.3]\,eV for the band structure used here. Furthermore, the highest photon energies are found at the peaks of the generating field, but the three-step model predicts them near the zero crossings. It is remarkable that the instantaneous response model fits very well for low intensity, but at higher intensity both the $P^{\mathrm{(NL)}}$ and $J$ exhibit dents at the field crests. These dents are clearly visible by comparing with the simplified calculation that neglects the motion of the Bloch electrons (Extended Data Fig.\,\ref{figSpecPnm}). These dents are reminiscent of the current generated by a single Bloch electron in Fig.\,\ref{figSing}. This strengthens the interpretation that Bloch electron motion can be regarded as a mechanism of harmonic generation not only for real electrons, which are the origin of $J$ and which was considered from the beginning as a possible mechanism for HHG, but also for virtual electrons (coherences), which are the origin of $P^{\mathrm{(NL)}}$. 
In contrast to the original strong-field regime ($\gamma < 1$), where optical components are very easily damaged, the almost-strong-field regime is often reached in high-power lasers and other optical instruments. The results of this work, especially numerical pulse propagation at feasible computation times, will be useful for the design of such instruments.

\bibliography{Bloch_Bib}


\section*{Methods}\label{sec11}

\subsection*{Calculations based on SBEs}

The consistent treatment of the OKE requires at least three bands \cite{RN249}. Two valence bands (bands 1 and 3) and one conduction band (band 2) are considered here. In a realistic band structure, valence bands have typically a transition energy on the order of 1\,eV, but resonance effects with the Vis-IR pulse do not prevail because many valence bands exist. To avoid resonance effects for only two valence bands, $\omega_{3}^k = \omega_{1}^k$ is assumed. However, only band 1 is coupled to the conduction band. Only the conduction band energy is considered to be $k$-dependent with a tight-binding band shape: 
\begin{equation}
\omega_2^k = \frac{1}{2} b_1 (1 - \mathrm{cos}(ka))
\label{eq:tightbinding} 
\end{equation}
with bandwidth $b_1 = 3$\,eV and lattice constant $a = 0.49$\,nm. The bandgap is set to $\omega_{12}^{k=0} = \omega_{12} = 10.3$\,eV, 

The dipole matrix elements are matched to the OKE at low intensities as described in Ref.\,\cite{RN249}. The valence band transitions are implemented with $d_{13}^k = d_{31}^k = d_{13} = 16$. The valence to conduction band transitions $d_{12}^{k} = d_{21}^{k}$ are implemented as \cite{RN218}
\begin{equation}
d_{12}^k = d_{12}^{k=0} \frac{\omega_{12}^{k=0}}{\omega_{12}^k}
. \label{DipoleElement} 
\end{equation}
with $d_{12}^{k=0} = 0.02$. All other dipole matrix elements are set to zero.

Relaxation is implemented as phenomenological damping terms. For the diagonal elements, the lifetime $T_1$ and the collision time $T_c$ are considered.
\begin{align}
\left( \partial_t \rho_{nn}^k  \right)_{relax}
= - \frac{1}{T_1} \rho_{nn}^{k} - \frac{1}{2 T_c} \left( \rho_{nn}^{k} - \rho_{nn}^{-k} \right)
. \label{T1} 
\end{align}
The lifetime in conduction bands of dielectrics usually exceeds 100\,fs, justifying the assumption $T_1 = \infty$. The terms proportional to $1/T_c$ causes a decay of the currents, while the total band population is preserved. The damping of the currents cannot be neglected, because Drude collision times are in the few-femtosecond range. This is in accordance with the qualitative picture of excited electrons that first undergo rapid momentum relaxation and thereafter energy relaxation on a longer timescale \cite{RN214}. For the coherences,  
\begin{equation}
\left( \partial_t \rho_{nm}^k  \right)_{relax}
= - \frac{1}{T_2} \rho_{nm}^{k}
. \label{T2} 
\end{equation}
where $T_2$ is the interband dephasing time. Here it is assumed that $T_2$ is identical for all coherences and independent on $k$. The relation between the dephasing time $T_2$ and the Drude collision time $T_c$ is not known. Here, $T_2 = 2 T_c$ is assumed, following the phenomenological picture that if scattering occurs to an electron at position $k$, its interband- and intraband coherences are likewise destroyed. 

\begin{figure}[h]%
\centering
\includegraphics[width=1\textwidth]{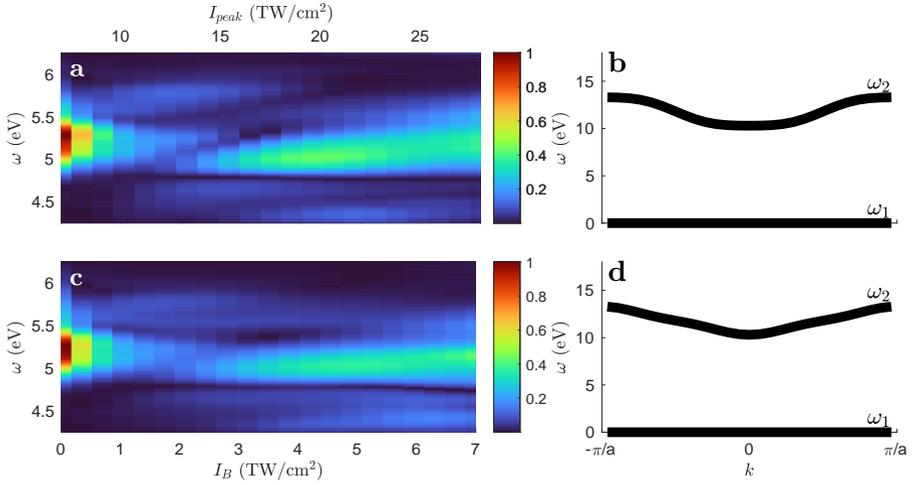}
\caption{\textbf{(Extended Data) The influence of the band shape}. \textbf{a} and \textbf{c} are the same as Fig.\,\ref{figProp}\textbf{e}, except that the shape of the conduction band (\ref{eq:tightbinding}) has been replaced by $\omega_2^k = \frac{1}{2} b_1 (1 - \mathrm{cos}(ka) - b_3 \left( \mathrm{cos}(3ka) - \mathrm{cos}(ka) \right) )$ with $b_3=-0.1$ (\textbf{a}) and $b_3=0.1$ (\textbf{c}). The band shape is displayed in \textbf{b} for $b_3=-0.1$ and in \textbf{d} for $b_3=0.1$.
}
\label{figBand3}
\end{figure}

\begin{figure}[h]%
\centering
\includegraphics[width=1\textwidth]{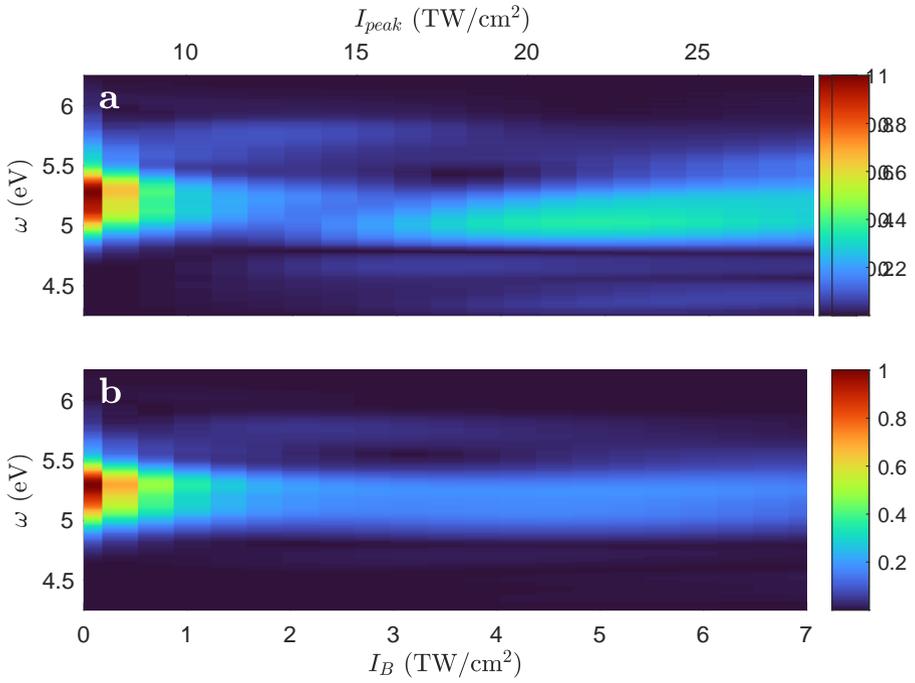}
\caption{\textbf{(Extended Data) Comparison of dephasing times}. The same as Fig.\,\ref{figProp}\textbf{e}, but using $T_2 = 10$\,fs (\textbf{a}) and $T_2 = 3$\,fs (\textbf{b}).}
\label{figTdeph}
\end{figure}

The numerical calculations are perfomed on a $k$-grid with 27 points. The time-domain integration is performed using the 4th-order Runge–Kutta (RK4) method. For the calculations with pulse propagation (Figs.\,\ref{figProp}, \ref{figBand3} and \ref{figTdeph}), a $t$-grid with 30001 points in the interval [-250, 250]\,fs is used. For the calculations without pulse propagation (all other Figures), a $t$-grid with 70001 points in the interval [-500, 500]\,fs is used.

\subsection*{Iteration of SBEs}

For dielectrics, the population transfer into the conduction band is only a small fraction of the valence band population when irreversible material changes are avoided. This justifies the assumptions $\rho_{11}^k -\rho_{22}^k = 1$; $\rho_{11}^k -\rho_{33}^k = 0$; $\rho_{33}^k -\rho_{22}^k = 1$, which effectively decouples the diagonal and off-diagonal elements of the SBEs. 
With this approximation the non-diagonal elements of (\ref{VelGauge}) can be transformed to
\begin{align}
\mathcal{F} \left\{ \rho_{12}^{k + A} \right\}
&= \frac{\mathcal{F} \left\{ {d}_{12}^{k + A} E \right\} + D_{12}}{\omega_{12} -\omega + i/T_2} \nonumber
\\
\mathcal{F} \left\{ \rho_{32}^{k + A}  \right\}
&= \frac{D_{32}}{\omega_{32} - \omega + i/T_2} \nonumber
\\
\mathcal{F} \left\{ \rho_{13}^{k + A}  \right\}
&= \frac{D_{13}}{\omega_{13} - \omega + i/T_2}
\label{DiagsFS} 
\end{align}
with
\begin{align}
D_{12} &= \mathcal{F} \left\{  -\widetilde{\omega}_{12}^{k + A} \rho_{12}^{k + A}  \right\}
-{d}_{13} \mathcal{F} \left\{ E \rho_{32}^{k + A} \right\} \nonumber
\\
D_{32} &= \mathcal{F} \left\{  -\widetilde{\omega}_{32}^{k + A} \rho_{32}^{k + A}  \right\}
-{d}_{13} \mathcal{F} \left\{ E \rho_{12}^{k + A} \right\} \nonumber
\\
D_{13} &= 
- \left. \mathcal{F} \left\{ d_{12}^{k + A} E  \rho_{32}^{k + A} \right\}^\ast \right\vert _{-\omega}
. \label{DnmFS} 
\end{align}
Here, the time-dependent transition energy $\omega_{nm}^{k + A}$ is separated into a static part $\omega_{nm} = \omega_{nm}^k$ and the dynamic part $\widetilde{\omega}_{nm}^{k + A} = \omega_{nm}^{k + A} - \omega_{nm}$. The elements $D_{nm}$ are the corrections due to the nonlinearity. A recursive method is used for their calculation: In the $n^{th}$ step of iteration, the elements $\rho_{nm}$ in (\ref{DiagsFS}) are calculated using the elements $D_{nm}$ of the $(n-1)^{th}$ step. Each step of iteration requires (inverse) Fourier transforms and time-domain multiplications.  $D_{nm} = 0$ is assumed in step 0. 

The iteration is equivalent to a power series expansion. Accordingly there is an upper limit for the electric field above which the iteration does not converge. In the limit of a monochromatic field with frequency $\omega_0$, the convergence criterion is given by $\gamma > 1$ \cite{RN226}. 

The linear susceptibility follows from (\ref{DiagsFS}) by setting $D_{12}=0$ :
\begin{equation}
\chi^{(1)}
= \frac{\mathcal{F} \left\{ P  \right\}}{\mathcal{F} \left\{ E  \right\}}
= \sum_k \frac{2 \omega_{12}^{k} ({d}_{12}^{k})^2 }{(\omega_{12}^{k})^2 -\omega^2 + 1/T_2^2 + 2 i \omega / T_2} \delta k
\end{equation}

\subsection*{Pulse propagation}

Macroscopic pulse propagation is calculated using the UPPE 
\begin{equation}
\partial_z \hat{E} = 
i \left( \frac{\omega}{u} -K \right)\hat{E}  
- \frac{2 \pi \omega}{K c^2} \left(i \omega \hat{P}^{\mathrm{(NL)}} + \hat{J} \right),
\end{equation}
where the hat symbol indicates the Fourier transform in the dimensions of time and transverse space. In addition to the propagation direction $z$, one transverse dimension (the $x$-dimension) is included to account for the noncollinear geometry with $K = \sqrt{n_R^2 \frac{\omega^2}{c^2} - k_x^2}$. Numerical tables are used for the refractive index $n_R$, $c$ is the speed of light and $u$ is the group velocity of the Vis-IR pulse. The electric field is treated as scalar field, because all pulses are polarized perpendicular to the plane of incidence. 

The UPPE is integrated numerically using the split-step method with an $x$-grid with 81 points in the interval [-260, 260]\,$\mu$m and a $z$-grid with 401 points in the interval [0, 100]\,$\mu$m. 

Subsequent to the propagation inside the crystal, the light propagating collinearly to A with emission angle $-\frac{\alpha}{2}$ is calculated by
\begin{equation}
E(\omega) = \hat{E}(\omega, k_x)
\end{equation}
with $tan(-\frac{\alpha}{2}) = \frac{k_x}{\sqrt{(\omega / c)^2 + (k_x)^2}}$.

\section*{Acknowledgement}

This project was supported primarily by the Deutsche Forschungsgemeinschaft (DFG, German Research Foundation) 
via Priority Programme 1840 "Quantum Dynamics in Tailored Intense Fields (QUTIF)" (project ID 281272215) 
and
via project B1 in the Collaborative Research Centre 1375 "Nonlinear optics down to atomic scales (NOA)" (project ID 398816777).

\section*{Author Contributions}
JR conducted the experiment. ANP developed the theory. All authors contributed to the discussion and preparation of the manuscript.



\newpage

\newpage

\newpage

\newpage

\newpage

\newpage

\newpage

\newpage

\end{document}